\documentclass[journal]{IEEEtran}
\IEEEoverridecommandlockouts
\usepackage{booktabs} 
\usepackage{graphicx}
\usepackage{setspace}
\usepackage{amsmath}
\usepackage{dsfont}
\usepackage{amsfonts}
\usepackage{lipsum}
\usepackage{multicol}
\usepackage{float}
\usepackage{bm}
\usepackage{color}
\usepackage[dvipsnames]{xcolor}
\usepackage{etoolbox}
\usepackage{soul}
\usepackage{amssymb}
\usepackage[linesnumbered,ruled,vlined]{algorithm2e}
\usepackage{mathtools,amssymb}
\usepackage{caption}
\usepackage{float}
\usepackage{amsthm}
\usepackage[caption = false,subrefformat=parens,labelformat=parens]{subfig}
\usepackage{adjustbox}
\usepackage{afterpage}
\usepackage{multirow}
\usepackage{mathrsfs}

\def\BibTeX{{\rm B\kern-.05em{\sc i\kern-.025em b}\kern-.08em
    T\kern-.1667em\lower.7ex\hbox{E}\kern-.125emX}}
\begin{document}

\title{ROK Defense M\&S in the Age of Hyperscale AI:\\Concepts, Challenges, and Future Directions

\author{
Youngjoon Lee, Taehyun Park, Yeongjoon Kang, Jonghoe Kim, Joonhyuk Kang
}
}

\maketitle

\begin{abstract}
Integrating hyperscale AI into national defense M\&S(Modeling and Simulation), under the expanding IoMDT(Internet of Military Defense Things) framework, is crucial for boosting strategic and operational readiness.
We examine how IoMDT-driven hyperscale AI can provide high accuracy, speed, and the ability to simulate complex, interconnected battlefield scenarios in defense M\&S.
Countries like the United States and China are leading the adoption of these technologies, with varying levels of success.
However, realizing the full potential of hyperscale AI requires overcoming challenges such as closed networks, sparse or “long-tail” data, complex decision-making processes, and a shortage of experts.
Future directions highlight the need to adopt domestic foundation models, expand GPU/NPU investments, leverage large tech services, and employ open source solutions.
These efforts will enhance national security, maintain a competitive edge, and spur broader technological and economic growth.
With this blueprint, the Republic of Korea can strengthen its defense posture and stay ahead of emerging threats in modern warfare.

\end{abstract}

\section{Introduction}

In recent years, integrating AI technologies into defense M\&S(Modeling and Simulation) is crucial as global security threats evolve\cite{fawkes2017developments, nacouzi2024artificial}. 
The ability to simulate complex scenarios is crucial, and AI can revolutionize defense M\&S with its advanced algorithms \cite{davis2022artificial}. These systems provide unprecedented accuracy and speed, leading to better strategic decisions. 
Improved outcomes ensure greater national security, making it essential to stay ahead of adversaries.
IoMDT(Internet of Military Defense Things)-equipped AI-driven models precisely predict enemy movements by fusing data from interconnected military IoT assets, while accelerated simulations allow near real-time decisions across heterogeneous platforms \cite{mcdowell2024re}.
Increased computational power supports complex analyses, while data-driven insights improve tactical responses \cite{10638797}. 
In addition, AI can efficiently streamline resource allocation \cite{morgan2020military}.

Countries around the world are recognizing the strategic importance of AI in defense. 
Nations such as United States (U.S.), China, Japan, European Union, and United Kingdom have comprehensive AI strategies, which highlight the role of AI in modern warfare.
Specifically, these nations integrate IoT-based defense architectures to better coordinate troops, optimize supply lines, and automate early warnings.
The Republic of Korea (ROK) has also announced defense AI strategies to adopt AI to remain competitive.
A key part of these strategies involves leveraging IoMDT to optimize force deployment, logistics, and battlefield command through interconnected devices.
National defense strategies continue to advance at an unprecedented pace and AI integration is now a global priority. 
Delayed implementation of AI technologies can have severe consequences for military capabilities, particularly given the rapid advancement of AI applications in defense systems.
Therefore, early adopters will gain significant advantages, with investment in AI technologies increasing worldwide \cite{madison2024scalable}.

\begin{figure}
     \centering
     \includegraphics[width=1.0\columnwidth]{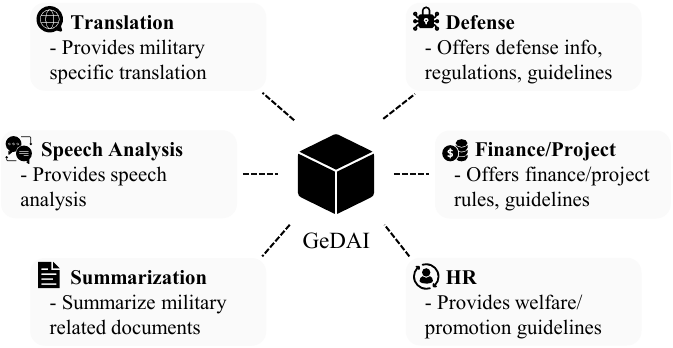}
     \caption{Illustration of an example of Generative AI used by the Republic of Korea Ministry of National Defense.}
     \label{fig:0}
\end{figure}

Recently, many countries have taken advantage of AI and IoMDT technologies to address previously unexplored areas in traditional defense M\&S, ranging from soldier health-monitoring wearables to autonomous surveillance systems \cite{caballero2024large}. 
This integration not only solves uncharted challenges, but also automates tasks that previously required significant manual effort and time.
For example, the ROK Ministry of National Defense (MND) has developed and is using an LLM called GeDAI (Generative Defense AI), as shown in Fig. \ref{fig:0}, to streamline various processes.
Specifically, GeDAI function like ChatGPT, enabling the ROK military to easily access the information they need.
This approach demonstrates a shift towards more efficient, AI-driven solutions that enhance operational capabilities and reduce reliance on human intervention in critical defense tasks.
By pairing IoMDT-based data with large-scale generative or predictive models enables real-time, data-driven analysis and rapid field adaptation, moving beyond traditional simulations.
Furthermore, the implementation of AI technologies such as GeDAI is expected to strengthen both the efficiency of decision-making processes and the effectiveness of strategic planning within the defense sectors. 

Main contributions are as follows:
\begin{itemize}
    \item Analysis of the differences in Defense M\&S operation methods and AI adoption strategies between U.S. and ROK. This comparison highlights key areas where both nations diverge in their approach to modernizing defense systems with AI.
    \item Identification of the challenges in integrating AI into Defense M\&S in ROK. These challenges include technical, operational, and policy characteristics and requirements specific to ROK's defense environment.
    \item Proposal of AI adoption strategies tailored not only to U.S. strategies, but also specialized for ROK. The strategies are specifically designed to take into account the current environment of ROK military.
\end{itemize}

The remainder of this paper is organized as follows. In section \ref{sec:concepts}, we explore key concepts of defense M\&S and hyperscale AI. In section \ref{sec:challenges}, the challenges of applying AI into ROK defense M\&S are described. Subsequently, future directions are presented in section \ref{sec:future}. Finally, we present our concluding remarks in section \ref{sec:conclusion}.

\section{Concepts}\label{sec:concepts}
In this section, we introduce the fundamental concepts of defense M\&S and hyperscale AI, as well as their relationship within an IoMDT environment.
\subsection{Defense M\&S}
\begin{figure}[h]
     \centering
     \includegraphics[width=1.0\columnwidth]{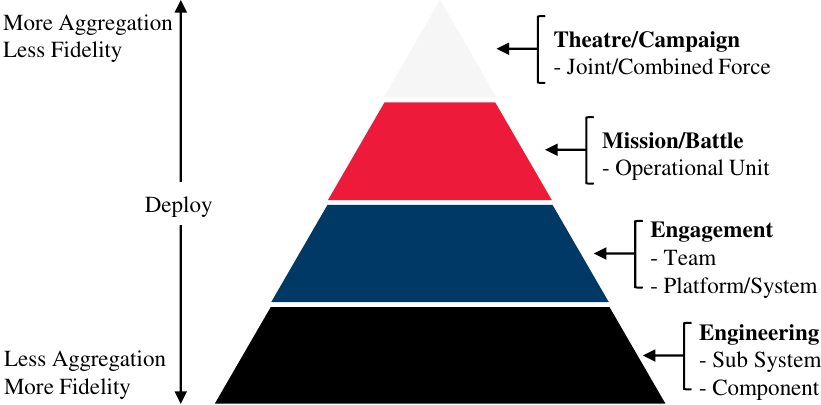}
     \caption{Illustration of defense M\&S hierarchical structure.}
     \label{fig:1}
\end{figure}

Defense M\&S is a critical tool that enables military organizations to create virtual representations of real-world systems, processes, and scenarios \cite{pace2004modeling, national2006defense}. 
By simulating combat situations, training exercises, IoMDT device inter connectivity, and equipment performance, defense M\&S allows for thorough analysis and preparation without the risks and costs associated with live experiments.
As shown in Fig. \ref{fig:1}, the hierarchical levels of simulation range from constructive simulations at the base, through virtual simulations in the middle, to live simulations at the apex. 
For example, using defense M\&S to predict the outcomes of battle scenarios can significantly reduce the expenses of conducting full-scale field exercises, as potential issues can be identified and addressed within the simulation environment.

As shown in the ROK part of Table \ref{tab:table0}, the ROK's M\&S operation plan involves a multilayered structure with various institutions sharing responsibilities. 
The ROK MND leads the field in policy development, budget management, and international cooperation. 
Specialized centers like the Joint Analysis Center and Joint Battle Simulation Center focus on requirements analysis, system operation, and ensuring interoperability. 
Each military branch establishes its own M\&S policies, maintains systems, and trains specialists. 
Agencies such as KIDA, DAPA, and ADD contribute to technology development and strategic analysis, emphasizing a decentralized approach to managing M\&S activities.

In contrast, the U.S. centralizes its M\&S operations under the Department of Defense (DoD), as illustrated in the U.S. part in Table \ref{tab:table0}. 
Key roles are played by offices like USD AT\&L and the DoD M\&S Coordination Office, which focus on strategic planning, policy approval, and coordination across training, intelligence, and analysis communities. 
Both countries emphasize advancing M\&S capabilities through coordinated efforts and strategic planning. 
Commonalities include a focus on policy development, interoperability, international cooperation, and the management of M\&S standards and architectures. 
A notable difference lies in organizational structure: ROK employs multiple agencies handling specific M\&S aspects, while the U.S. adopts a centralized model with the DoD M\&S Coordination Office serving as a focal point. 

\subsection{Hyperscale AI}
\begin{figure}[h]
     \centering
     \includegraphics[width=\columnwidth]{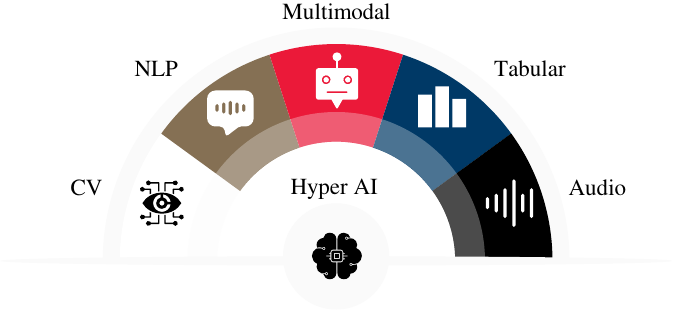}
     \caption{Illustration of fields where hyperscale AI can be applied.}
     \label{fig:3}
\end{figure}

Hyperscale AI, when paired with IoMDT sensor data, operates on an unprecedented scale, leveraging massive computational resources and diverse battlefield datasets to solve complex operational problems \cite{zhou2023comprehensive}. 
As illustrated in Fig. \ref{fig:3}, possible applications of hyperscale AI include compter vision, natural language, multimodal, tabular, and audio processing. 
Hyperscale AI is being used to enhance predictive analytics, automate decision-making processes, and provide deep insights through machine learning algorithms. 
For instance, applying hyperscale AI to defense M\&S, it is possible to conduct wargames with reduced human intervention as COA-GPT \cite{goecks2024coa}.
This allows simulation of numerous scenarios, strategic decision making, and outcome predictions, thereby decreasing the time and resources required for human-led simulations.

\begin{table}[h]
\caption{Representative foundation models announced by Korean and American companies (2023-2024).}
\label{tab:table1}
\resizebox{\columnwidth}{!}{%
\begin{tabular}{cllll}
\hline
\textbf{Country}      & \textbf{Date} & \textbf{Model}      & \textbf{Type}          & \textbf{Creator(s)} \\ \hline
\multirow{11}{*}{ROK}  & July 2023     & Exaone 2.0          & Large Multimodal Model & LG                  \\
 & Aug. 2023 & HyperCLOVA X      & Large Language Model   & Naver             \\
 & Aug. 2023 & VARCO             & Large Language Model   & NCSoft            \\
 & Oct. 2023 & Mi:dm             & Large Multimodal Model & KT                \\
                       & Nov. 2023     & Samsung Gauss       & Large Multimodal Model & Samsung Electronics \\
 & Feb. 2024 & Solar             & Large Language Model   & Upstage           \\
 & Mar. 2024 & Karlo 2.1         & Text-to-Image Model    & Kakao             \\
 & Mar. 2024 & Marengo 2.6       & Large Multimodal Model & Twelve Labs       \\
 & Apr. 2024 & KoGPT 2.0         & Large Language Model   & Kakao             \\
 & Apr. 2024 & DASH              & Large Language Model   & Naver             \\
 & Jun. 2024 & Telco             & Large Language Model   & SKT               \\
 & Jul. 2024 & GeDAI             & Large Language Model   & ROK MND           \\
 & Nov. 2024 & Solar Pro            & Large Language Model   & Upstage           \\
 & Dec. 2024 & EXAONE 3.5             & Large Language Model   & LG AI Research               \\ \hline

\multirow{16}{*}{U.S.} & Mar. 2023     & Stable Diffusion v2 & Text-to-Image Model    & Stability AI        \\
 &Mar. 2023 & Claude & Large Language Model & Anthropic \\
 & Apr. 2023 & Segment Anything  & Image Segmentation     & Meta              \\
 & Jul. 2023 & Llama 2           & Large Language Model   & Meta              \\
 & Aug. 2023 & DALL-E 3          & Image Generation       & Microsoft, OpenAI \\
 & Aug. 2023 & SynthID           & Watermarking           & Google, DeepMind  \\
 & Nov. 2023 & GPT-4 Turbo       & Large Language Model   & Microsoft, OpenAI \\
 &Nov. 2023 & Claude 2.1 & Large Language Model & Anthropic \\
 &Nov. 2023 & Grok-1 & Large Language Model & xAI \\
 & Nov. 2023 & Whisper v3        & Speech-to-Text         & Microsoft, OpenAI \\
 & Nov. 2023 & Claude 2.1        & Large Language Model   & Anthropic         \\
 & Nov. 2023 & Inflection-2      & Large Language Model   & Inflection        \\
 & Dec. 2023 & Gemini            & Large Language Model   & Google            \\
 & Dec. 2023 & Midjourney v6     & Text-to-Image Model    & Midjourney        \\
 & Mar. 2024 & Claude 3.0 & Large Language Model & Anthropic \\
 & Apr. 2024 & Llama 3           & Large Language Model   & Meta              \\
 & May 2024 & GPT-4o & Large Language Model & Microsoft, OpenAI \\
 & May 2024 & Chameleon & Large Multimodal Model   & Meta         \\
 & Jun. 2024 & Claude 3.5 Sonnet & Large Language Model   & Anthropic         \\
 & Jul. 2024 & GPT-4o mini       & Large Language Model   & Microsoft, OpenAI \\
 & Aug. 2024 & Grok-2 & Large Language Model & xAI \\
 & Sep. 2024 & LLaMA 3.2 & Large Language Model & Meta \\
 & Dec. 2024 & LLaMA 3.3 & Large Language Model & Meta \\
 & Dec. 2024 & Gemini 2.0 & Large Language Model & Google \\\hline
\end{tabular}%
}
\end{table}

As shown in Table \ref{tab:table1}, ROK is consistently releasing diverse foundation models, much like U.S. companies.
These models range from large language models to multimodal and text-to-image models, showcasing ROK's active participation in advancing AI technologies.
A notable difference between the two countries is the accessibility of these foundation models.
Unlike ROK, U.S. is actively trying to adopt services created by its big tech companies, integrating AI solutions from firms like OpenAI, Google, and Meta into various industries. 
This strategy accelerates innovation while expanding the influence of AI technologies across various sectors.

\begin{figure}[t]
\centering
\includegraphics[width=\columnwidth]{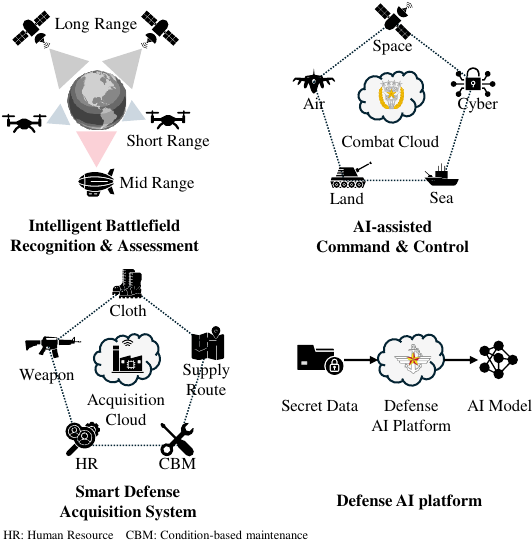}
\caption{Comprehensive illustration of ROK's IoMDT architecture.}
\label{fig:2}
\end{figure}

The ROK military is focusing on four key domains for IoMDT-driven AI applications, as illustrated in Fig. \ref{fig:2}.
Within these domains, ROK can develop specific AI capabilities tailored to military needs.
By leveraging hyperscale AI technologies from domestic tech companies like NAVER, Kakao, and Samsung Electronics, these military domains can be significantly enhanced.
The advanced language processing and multimodal capabilities of commercial AI models could improve battlefield analysis and decision-making processes by processing diverse IoMDT data streams in real-time.
Such integration would enhance both operational efficiency and combat effectiveness while maintaining security through the military's specialized IoMDT platforms.

\subsection{IoMDT for Next-Generation Simulations}
While commercial IoT systems focus on consumer-centric devices, IoMDT centers on mission-critical devices, sensors, and applications specialized for military operations. 
By merging IoMDT data with hyperscale AI capabilities, defense M\&S can capture a more comprehensive digital twin of the battlefield.
For instance, real-time status from soldier-worn biosensors, UAV imagery, radar signals, and supply chain IoT nodes feed directly into high-fidelity simulations.
This integrated architecture offers an adaptive loop: sensor data update simulations, while simulation outcomes inform AI-driven optimization or predictive analytics for resource allocation, mission planning, and threat detection.
Moreover, IoMDT frameworks require stringent cybersecurity and reliability measures, distinct from commercial IoT. 
The robust encryption, secure authentication, and redundant network topologies must ensure that IoMDT data seamlessly reaches the M\&S pipeline without compromising battlefield security.

\begin{table*}
\caption{Division of responsibilities among institutions related to defense M\&S in Republic of Korea and United States.}
\label{tab:table0}
\resizebox{\textwidth}{!}{%
\begin{tabular}{cll}
\hline
\textbf{Country} &
  \textbf{Institution (Division)} &
  \textbf{Mission} \\ \hline
\multirow{13}{*}{ROK} &
  ROK MND &
  \begin{tabular}[c]{@{}l@{}}• Develop and coordinate defense M\&S policies\\ • Review defense M\&S system requirements\\ • Manage defense M\&S budget\\ • Run Defense CIO Working Group and M\&S Council\\ • Oversee international M\&S cooperation\end{tabular} \\ \cline{2-3} 
 &
  Joint Analysis Center &
  \begin{tabular}[c]{@{}l@{}}• Analyze defense M\&S requirements\\ • Coordinate joint experiments\\ • Update JCS models\\ • Standardize M\&S data\end{tabular} \\ \cline{2-3} 
 &
  Joint Battle Simulation Center &
  \begin{tabular}[c]{@{}l@{}}• Operate M\&S systems\\ • Ensure M\&S system interconnectivity\end{tabular} \\ \cline{2-3} 
 &
  Force Planning Directorate &
  • Review defense M\&S requirements \\ \cline{2-3} 
 &
  C4I Directorate &
  • Support M\&S system interoperability \\ \cline{2-3} 
 &
  Each Military Branch &
  \begin{tabular}[c]{@{}l@{}}• Establish branch-specific M\&S policies\\ • Develop M\&S budget requests\\ • Maintain branch M\&S systems\\ • Train M\&S specialists\end{tabular} \\ \cline{2-3} 
 &
  USFK (United States Forces Korea) &
  \begin{tabular}[c]{@{}l@{}}• Plan US-Korea joint simulations\\ • Ensure M\&S system interconnectivity \\ • Manage Korean War Game System maintenance\end{tabular} \\ \cline{2-3} 
 &
  \begin{tabular}[c]{@{}l@{}}DAPA\\ (Defense Acquisition Program Administration)\end{tabular} &
  \begin{tabular}[c]{@{}l@{}}• Manage M\&S acquisition policies\\ • Plan and assess M\&S technology needs\\ • Oversee defense M\&S projects\end{tabular} \\ \cline{2-3} 
 &
  M\&S Research Group &
  \begin{tabular}[c]{@{}l@{}}• Develop defense M\&S core technologies\\ • Support institute-managed M\&S projects\end{tabular} \\ \cline{2-3} 
 &
  Defense Technology Support Center &
  \begin{tabular}[c]{@{}l@{}}• Support company M\&S projects\\ • Manage core tech development projects\end{tabular} \\ \cline{2-3} 
 &
  \begin{tabular}[c]{@{}l@{}}KIDA\\ (Korea Institute of Defense Analyses)\end{tabular} &
  \begin{tabular}[c]{@{}l@{}}• Develop branch-specific M\&S policies\\ • Manage international M\&S cooperation\\ • Research advanced M\&S methods\\ • Analyze defense strategies and experiments\end{tabular} \\ \cline{2-3} 
 &
  \begin{tabular}[c]{@{}l@{}}DTaQ\\ (Defense Agency for Technology and Quality)\end{tabular} &
  \begin{tabular}[c]{@{}l@{}}• Support M\&S technology planning\\ • Research M\&S systems \\ • Conduct HLA tests\end{tabular} \\ \cline{2-3} 
 &
  \begin{tabular}[c]{@{}l@{}}ROK DCC\\ (Defense Communication Command)\end{tabular} &
  \begin{tabular}[c]{@{}l@{}}• Evaluate M\&S interoperability\\ • Manage M\&S standards and architectures\end{tabular} \\ \cline{1-2}
\multirow{8}{*}{U.S.} &
  \begin{tabular}[c]{@{}l@{}}USD AT\&L\\ (Undersecretary of Defense for Acquisition, Technology and Logistics)\end{tabular} &
  \begin{tabular}[c]{@{}l@{}}• Coordinates development of DoD M\&S metadata search\\ • Approves, publishes M\&S strategic plans and reports\\ • Establishes and chairs DoD M\&S steering committee\\ • Represents DoD Test \& Evaluation on M\&S committee\\ • Represents DoD Acquisition on M\&S committee\\ • Develops M\&S strategic plans for Acquisition/Evaluation communities\end{tabular} \\ \cline{2-3} 
 &
  \begin{tabular}[c]{@{}l@{}}DoD M\&SCO\\ (Modeling and Simulation Coordination Office)\end{tabular} &
  \begin{tabular}[c]{@{}l@{}}• Serves as DoD focal point for M\&S coordination\\ • Provides oversight for M\&S projects/activities\\ • Manages USD (AT\&L) M\&S plans and investments\\ • Coordinates international M\&S activities for DoD\\ • Chairs M\&S working groups as directed\\ • Publishes DoD M\&S glossary and manages changes\end{tabular} \\ \cline{2-3} 
 &
  \begin{tabular}[c]{@{}l@{}}USD P\&R\\ (Under Secretary of Defense for Personnel and Readiness)\end{tabular} &
  \begin{tabular}[c]{@{}l@{}}• Represent Training community on the M\&S committee\\ • Develop Training community M\&S strategic plan\end{tabular} \\ \cline{2-3} 
 &
  \begin{tabular}[c]{@{}l@{}}USD I\&S\\ (Under Secretary of Defense for Intelligence and Security)\end{tabular} &
  \begin{tabular}[c]{@{}l@{}}• Represent Intelligence community on the M\&S committee\\ • Develop Intelligence community M\&S strategic plan\end{tabular} \\ \cline{2-3} 
 &
  \begin{tabular}[c]{@{}l@{}}DCAPE\\ (Director, Cost Assessment and Program Evaluation)\end{tabular} &
  \begin{tabular}[c]{@{}l@{}}• Represent Analysis community on the M\&S committee\\ • Develop Analysis community M\&S strategic plan\end{tabular} \\ \cline{2-3} 
 &
  DoD Component Heads &
  \begin{tabular}[c]{@{}l@{}}• Develop metadata for key M\&S assets\\ • Review metadata for reuse before development\\ • Advise USD (AT\&L) on M\&S capabilities/uses\end{tabular} \\ \cline{2-3} 
 &
  Each Military Branch &
  • Provide each representative to the M\&S committee \\ \cline{2-3} 
 &
  \begin{tabular}[c]{@{}l@{}}CJCS\\ (Chairman of the Joint Chiefs of Staff)\end{tabular} &
  \begin{tabular}[c]{@{}l@{}}• Represent Combatant Commands on the M\&S committee\\ • Represent Planning and Experimentation communities\\ • Develop M\&S strategic plans\end{tabular} \\ \hline
\end{tabular}%
}
\end{table*}

\section{Core Challenges}\label{sec:challenges}
In this section, we identify and discuss the principal obstacles that arise when integrating AI into ROK defense M\&S.
\subsection{Closed Network}
One of the major challenges in adopting IoMDT-enhanced AI in defense M\&S is operating within closed networks, where the secure integration of interconnected devices is restricted.
In practice, military secrets are hard to utilize for AI training in a closed network environment.
However, open source tools and libraries are foundational for modern AI development, offering cutting-edge algorithms and efficient frameworks that accelerate innovation. 
Without access to these resources, researchers are compelled to invest time and effort in developing basic functionalities from scratch, leading to redundant work and slower progress. 
This not only delays project timelines, but also increases costs, as proprietary solutions require continuous maintenance and updates. 
Furthermore, limited interaction with the open source community isolates defense researchers from collaborative progress, impeding information sharing and the adoption of best practices.

Similarly, the inability to leverage publicly available datasets poses a critical challenge for AI development in defense M\&S. 
High-quality, large-scale datasets are essential for training robust AI models, particularly in machine learning and deep learning applications. 
Without access to such data, researchers are forced to depend on restricted or synthetic datasets that may not adequately represent real-world scenarios.
This limitation can result in models with poor generalization capabilities, reducing their effectiveness in practical applications. 
Furthermore, generating proprietary datasets is often resource-intensive and may not match the diversity and scale of public datasets, thereby impeding the development of advanced AI solutions within the defense sector.

\subsection{Long-Tail Data}
Long-tail data challenges significantly hinder IoMDT-driven AI development in defense M\&S, particularly when sensor inputs and operational logs from rare or adversarial scenarios are limited or unavailable.
For example, while extensive data may be available for ROK unmanned aerial vehicles (UAVs), there is a lack of data on enemy UAVs, limiting the AI's ability to accurately model or predict adversarial behavior. 
This absence impedes the development of robust AI models equipped to handle a broad spectrum of scenarios, particularly those involving adversarial equipment or tactics that remain insufficiently documented or inaccessible.
Consequently, AI systems may underperform in critical situations where understanding enemy actions is vital.

Moreover, acquiring data through actual combat experiments is prohibitively expensive and resource-intensive, often resulting in only a small subset of the necessary data being collected. 
The high costs associated with these experiments mean that researchers must operate with limited datasets, which may not capture the full spectrum of operational variables and conditions. 
This scarcity of comprehensive data can result in AI models that lack effectiveness or exhibit greater susceptibility to errors when applied to real-world scenarios, thus reducing the reliability and overall performance of defense M\&S applications.
Addressing this gap is crucial for developing AI solutions that can adapt to complex and unpredictable combat environments.

\subsection{Complex Decisions}
Importing external software or data into closed networks poses a major challenge for AI development in defense M\&S due to complex decision-making processes.
Researchers must navigate a labyrinth of approvals from numerous officials, each bearing responsibility for security and compliance. For instance, bringing in pretrained weights from publicly available AI models necessitates proving that they pose no risks of viruses or hacking. Should a security incident occur, the researchers are held accountable, which can have serious professional repercussions. 
This cumbersome process not only consumes valuable time but also cultivates an atmosphere of caution and risk aversion. 
Consequently, researchers may hesitate to explore innovative or experimental methodologies.

As a result, there is a tendency to rely solely on existing, familiar approaches, which can stifle innovation and slow the advancement of AI capabilities within the defense sector. 
The apprehension of potential consequences dissuades researchers from pursuing cutting-edge technologies that could provide considerable advancements.
This reliance on established methods may prevent the development of solutions that address emerging threats or capitalize on new opportunities. Addressing this challenge is crucial for fostering an environment where innovation thrives and researchers feel empowered to pursue advanced solutions without undue procedural obstacles.

\subsection{Scarcity of AI Experts}
Lastly, the scarcity of AI professionals poses a notable obstacle for defense M\&S in ROK, driven by the increasing wage disparity between large corporations and the military. Highly skilled AI developers and researchers are increasingly attracted to the private sector, where they receive more competitive compensation packages. This trend leads to a talent drain from the military and related institutions, hindering their ability to recruit and retain the expertise necessary for advanced AI development. The lack of sufficient incentives in the defense sector makes it difficult to attract top-tier talent, which is crucial for maintaining technological superiority and fostering innovation in defense applications.

Furthermore, the rotational position system for military officers in ROK exacerbates the shortage of specialized AI personnel. 
The officers are frequently reassigned to different roles, preventing them from acquiring sufficient expertise in the AI field.
This lack of continuity and specialization undermines the military's capacity to cultivate and retain skilled professionals who can drive AI initiatives forward.
The absence of dedicated AI specialists means that defense projects may suffer from a lack of technical proficiency, leading to suboptimal outcomes. 
Addressing this issue is essential for building a robust AI capability within the military, ensuring that personnel can develop and apply specialized knowledge effectively.

\section{Future Directions}\label{sec:future}
\begin{figure}[t]
     \centering
     \includegraphics[width=\columnwidth]{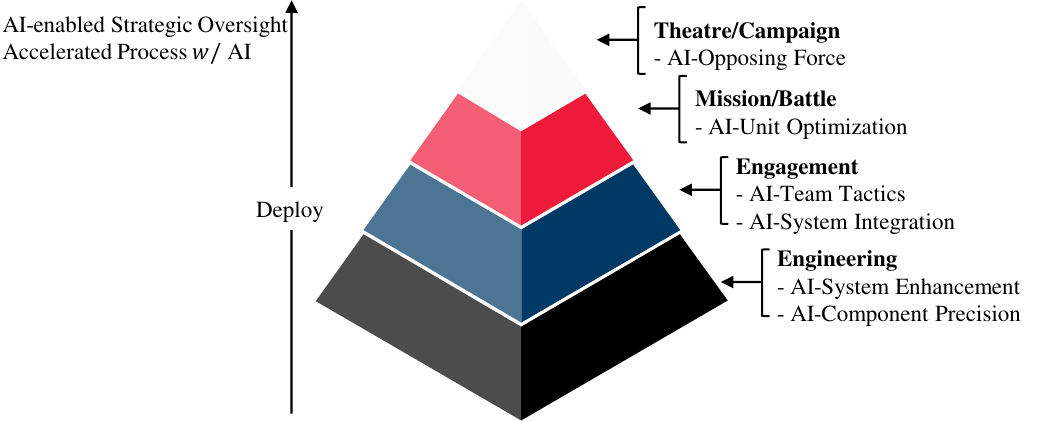}
     \caption{Illustration of defense M\&S with hyperscale AI.}
     \label{fig:4}
\end{figure}

In this section, we propose potential solutions and directions to achieve hyperscale AI-based defense M\&S in ROK as shown in Fig. \ref{fig:4}.

\subsection{Adopt Domestic Foundation Models}
A potential future direction for integrating AI into IoMDT-based defense M\&S in ROK is the active adoption of foundation models such as HyperCLOVA X and EXAONE 3.5 developed by domestic companies, mirroring strategies employed by U.S. DoD. 
By leveraging the expertise of private sector AI developers, the military can accelerate the implementation of advanced AI technologies. 
These foundation models, already robust and well-tested, can provide a solid starting point for defense applications, reducing development time and costs. 
Collaborating with Korean AI companies can also foster innovation and ensure that the models are tailored to the specific needs and contexts of ROK's defense landscape.

While transferring data to private companies raises concerns about security and potential leaks, this collaboration can help address the shortage of AI professionals within the military. 
By engaging with external experts, the defense sector can benefit from specialized knowledge and skills that may not be readily available internally. 
Furthermore, these partnerships can offer valuable insights into which data should be collected through combat experiments to optimize AI models for military use. 
By working closely with private companies, the military can develop a clearer direction for data acquisition, ensuring that AI models are effectively adapted to their operational requirements while mitigating security risks through strict data handling protocols.

\subsection{Involve Major IT Corporations}
Mirroring the strategy of the U.S. Defense Innovation Unit (DIU), a key future direction for ROK involves encouraging its leading IT corporations—such as NAVER, Samsung Electronics, and SKT to actively participate in R\&D initiatives within the defense M\&S sector.
Through strategic collaboration with these industry leaders who are pioneering the IoT and IoMDT markets, the military can leverage cutting-edge connectivity, sensor fusion, and cloud-edge AI technologies to enhance defense M\&S performance.
This collaboration enables the integration of advanced IoT solutions into wargame and operations, fostering innovation and bridging the gap between civilian technological advancements and defense applications. Such partnerships can accelerate the development of sophisticated simulation models and AI systems tailored to the military's specific needs.

Engaging large IT companies in defense projects also allows the military to rapidly acquire new weapon systems by adapting existing corporate solutions for military use. 
These companies possess cutting-edge technologies and specialized expertise that can be optimized to address defense requirements, thereby reducing both the development time and associated costs.
By collaborating with leading technology enterprises in the IoMDT sector, the military can ensure that its defense infrastructure is built using the most efficient and secure technologies available.
This not only enhances operational capabilities, but also addresses the shortage of specialized AI personnel by tapping into the skilled workforce of these corporations. Ultimately, this approach can strengthen national defense while fostering a synergistic relationship between the military and the private sector.

\subsection{Diversify GPU/NPU Infrastructure}
The need for a diverse GPU/NPU infrastructure is a critical future direction for integrating AI into IoMDT-based defense M\&S in ROK, ensuring robust on-site computing for the vast sensor networks deployed in modern battle environments. 
Currently, getting NVIDIA AI hardware is challenging, and over-reliance on a single company's products can lead to vulnerabilities where national defense capabilities might be influenced or constrained by corporate decisions. 
This dependency poses risks to security and operational readiness, as any disruptions in the supply chain or changes in corporate policies could adversely affect the military's AI capabilities. Therefore, it is imperative to reduce reliance on a single vendor to ensure robustness and autonomy in defense technology.

By adopting and utilizing GPUs and NPUs from multiple sources such as Gaudi series of Intel, the military can build a more resilient and flexible AI infrastructure. 
This diversification not only mitigates the risks associated with supply chain disruptions but also fosters competitive innovation among suppliers, potentially leading to better performance and cost savings. 
Incorporating hardware from domestic companies like Samsung Electronics also supports the national industry and can enhance security through closer collaboration. 
A multivendor approach ensures that the defense sector is not held hostage by any single company's technology, thereby strengthening national security and technological sovereignty.

\subsection{Leverage Open Source Software \& Data}
Enhancing AI-based M\&S for confidential military data requires leveraging both proprietary and open source software and data.
Integrating open source tools and datasets enables the development of robust AI models without compromising sensitive data, ensuring security while leveraging cutting-edge innovations.
This strategy facilitates the in-house development of AI solutions customized for military requirements, harnessing the worldwide AI community's collaborative progress.

Achieving this integration requires the elimination of complex approval procedures when using externally available software and data internally. 
Simplifying administrative procedures and fostering a supportive environment accelerates the adoption of new technologies in defense M\&S.
Reducing administrative hurdles promotes innovation and agility, keeping the military at the forefront of technological progress.
Cultivating a conducive atmosphere for technology adoption expedites development and draws talented individuals seeking a progressive and encouraging workplace.

\section{Conclusion}\label{sec:conclusion}
Integrating hyperscale AI and IoMDT into ROK’s defense M\&S offers significant opportunities and challenges, laying the groundwork for autonomous operations, enhanced situational awareness, and robust tactical decision support. Addressing obstacles such as closed networks, long-tail data scarcity, complex decision-making processes, and a shortage of AI experts is crucial for advancing the nation's defense capabilities.
Through the adoption of domestic foundation models, the participation of major IT corporations in defense R\&D, diversification of GPU/NPU infrastructure, and utilization of open source software and data, ROK can significantly enhance its defense M\&S capabilities.
These strategic directions address existing challenges while promoting innovation and operational readiness. 
This approach will enable ROK to strengthen national security and contribute to broader technological and economic growth in the age of hyperscale AI.

\section{Acknowledgment}
This research was supported by the MSIT (Ministry of Science and ICT), Korea, under the ITRC (Information Technology Research Center) support program (IITP-2025-2020-0-01787) supervised by the IITP (Institute of Information \& Communications Technology Planning \& Evaluation). Also, this research was supported in part by the NAVER-Intel Co-Lab. The work was conducted by KAIST and reviewed by both NAVER and Intel.

\bibliographystyle{IEEEtran}
\bibliography{reference}

\begin{thebibliography}{10}
\providecommand{\url}[1]{#1}
\csname url@samestyle\endcsname
\providecommand{\newblock}{\relax}
\providecommand{\bibinfo}[2]{#2}
\providecommand{\BIBentrySTDinterwordspacing}{\spaceskip=0pt\relax}
\providecommand{\BIBentryALTinterwordstretchfactor}{4}
\providecommand{\BIBentryALTinterwordspacing}{\spaceskip=\fontdimen2\font plus
\BIBentryALTinterwordstretchfactor\fontdimen3\font minus \fontdimen4\font\relax}
\providecommand{\BIBforeignlanguage}[2]{{%
\expandafter\ifx\csname l@#1\endcsname\relax
\typeout{** WARNING: IEEEtran.bst: No hyphenation pattern has been}%
\typeout{** loaded for the language `#1'. Using the pattern for}%
\typeout{** the default language instead.}%
\else
\language=\csname l@#1\endcsname
\fi
#2}}
\providecommand{\BIBdecl}{\relax}
\BIBdecl

\bibitem{fawkes2017developments}
A.~J. Fawkes, ``Developments in artificial intelligence: Opportunities and challenges for military modeling and simulation,'' in \emph{in Proc. NATO M\&S Symp.}, vol.~11, no. 1-12, 2017.

\bibitem{nacouzi2024artificial}
G.~Nacouzi, O.~A. Osoba, J.~Tran, and S.~Ishikawa, ``Artificial intelligence and machine learning applications for defensive counterspace,'' \emph{Santa Monica: RAND Corporation}, 2024.

\bibitem{davis2022artificial}
P.~K. Davis and P.~Bracken, ``Artificial intelligence for wargaming and modeling,'' \emph{J. Def. Model. Simul.}, p. 15485129211073126, 2022.

\bibitem{mcdowell2024re}
K.~McDowell, E.~Novoseller, A.~Madison, V.~G. Goecks, and C.~Kelshaw, ``Re-envisioning command and control,'' in \emph{in Proc. Int. Conf. Mil. Commun. Inf. Syst. (ICMCIS)}.\hskip 1em plus 0.5em minus 0.4em\relax IEEE, 2024, pp. 1--7.

\bibitem{10638797}
D.~H. Hagos and D.~B. Rawat, ``Neuro-symbolic ai for military applications,'' \emph{IEEE Trans. Artif. Intell.}, pp. 1--15, 2024.

\bibitem{morgan2020military}
F.~E. Morgan, B.~Boudreaux, A.~J. Lohn, M.~Ashby, C.~Curriden, K.~Klima, and D.~Grossman, ``Military applications of artificial intelligence,'' \emph{Santa Monica: RAND Corporation}, 2020.

\bibitem{madison2024scalable}
A.~Madison, E.~Novoseller, V.~G. Goecks, B.~T. Files, N.~Waytowich, A.~Yu, V.~J. Lawhern, S.~Thurman, C.~Kelshaw, and K.~McDowell, ``Scalable interactive machine learning for future command and control,'' in \emph{in Proc. Int. Conf. Mil. Commun. Inf. Syst. (ICMCIS)}.\hskip 1em plus 0.5em minus 0.4em\relax IEEE, 2024, pp. 1--10.

\bibitem{caballero2024large}
W.~N. Caballero and P.~R. Jenkins, ``On large language models in national security applications,'' \emph{arXiv preprint arXiv:2407.03453}, 2024.

\bibitem{pace2004modeling}
D.~K. Pace, ``Modeling and simulation verification and validation challenges,'' \emph{Johns Hopkins APL technical digest}, vol.~25, no.~2, pp. 163--172, 2004.

\bibitem{national2006defense}
N.~R. Council, D.~on~Engineering, P.~Sciences, B.~on~Mathematical~Sciences, T.~Applications, C.~on~Modeling, and S.~for Defense~Transformation, \emph{Defense modeling, simulation, and analysis: meeting the challenge}.\hskip 1em plus 0.5em minus 0.4em\relax National Academies Press, 2006.

\bibitem{zhou2023comprehensive}
C.~Zhou, Q.~Li, C.~Li, J.~Yu, Y.~Liu, G.~Wang, K.~Zhang, C.~Ji, Q.~Yan, L.~He \emph{et~al.}, ``A comprehensive survey on pretrained foundation models: A history from bert to chatgpt,'' \emph{arXiv preprint arXiv:2302.09419}, 2023.

\bibitem{goecks2024coa}
V.~G. Goecks and N.~Waytowich, ``Coa-gpt: Generative pre-trained transformers for accelerated course of action development in military operations,'' in \emph{in Proc. Int. Conf. Mil. Commun. Inf. Syst. (ICMCIS)}.\hskip 1em plus 0.5em minus 0.4em\relax IEEE, 2024, pp. 01--10.

\end{thebibliography}

\section{Biographies}

\begin{IEEEbiography}[{\includegraphics[width=1in,height=1.25in,clip,keepaspectratio]{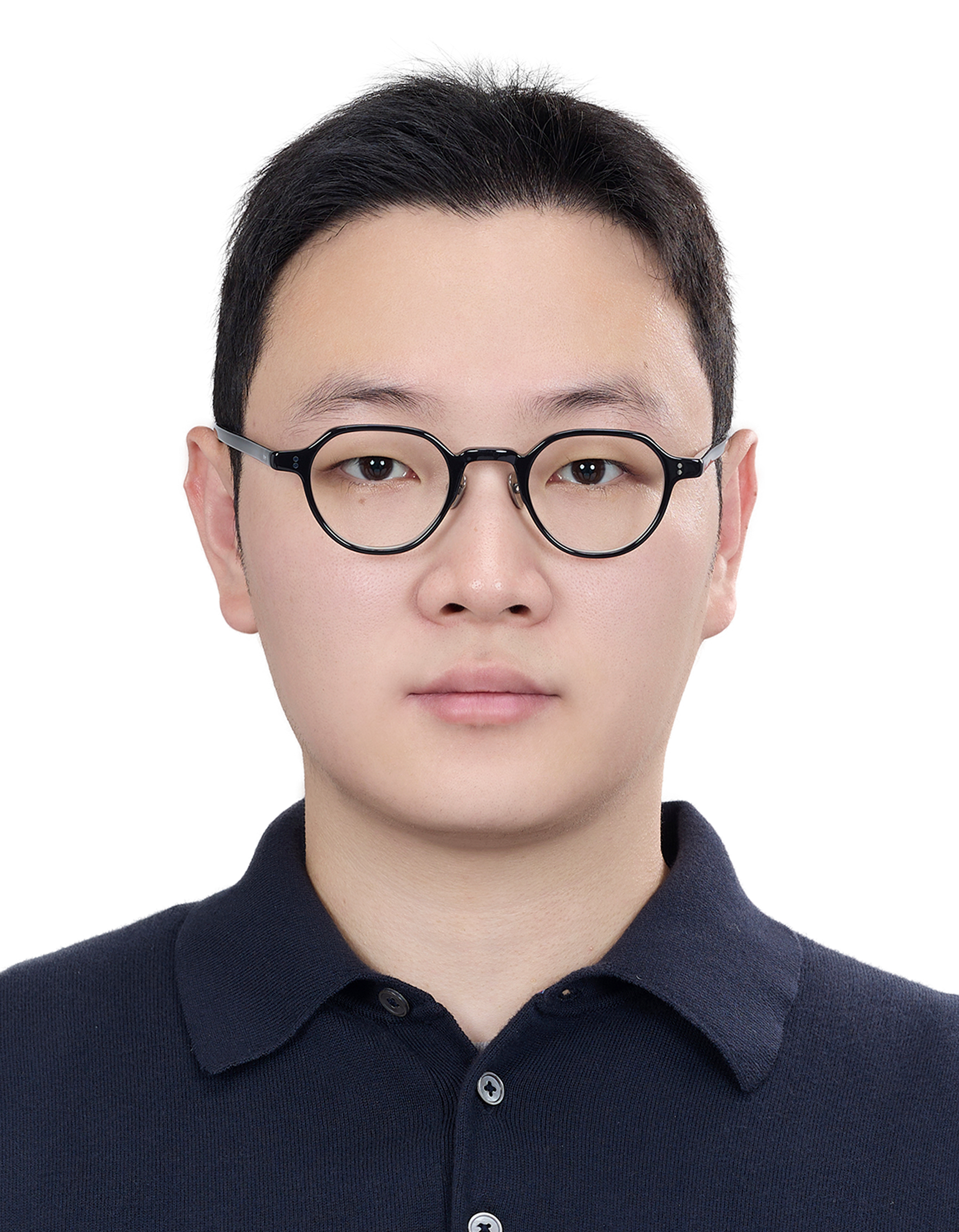}}]{Youngjoon Lee} \textcolor{black}{(yjlee22@kaist.ac.kr) received the B.S. degree in electrical engineering from UNIST, South Korea, in 2019, and the M.S. degree in electrical engineering from KAIST, South Korea, in 2021. From 2021 to 2023, he was a researcher with the Research Institute of Clean Manufacturing System, KITECH (Korea Institute of Industrial Technology), South Korea. He was a researcher with the Center for Military Analysis and Planning, KIDA (Korea Institute for Defense Analyses), South Korea, from 2023 to 2024. He is currently pursuing the Ph.D. degree under the supervision of Prof. Joonhyuk Kang at KAIST, South Korea. 
His research interests lie in practical, reliable AI and its application to privacy-preserving systems, as well as federated learning and synthetic data generation.
}
\end{IEEEbiography}
\vspace{-1cm}
\begin{IEEEbiography}[{\includegraphics[width=1in,height=1.25in,clip,keepaspectratio]{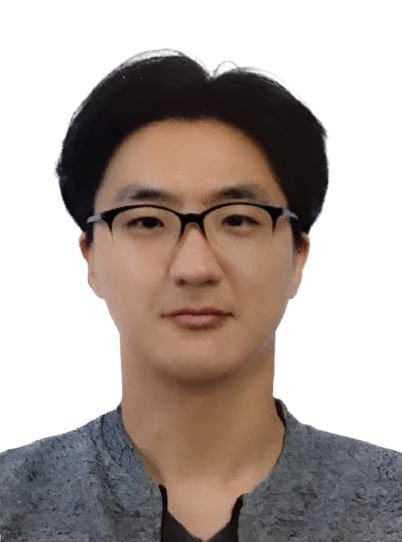}}]{Taehyun Park} \textcolor{black}{(tpark@kida.re.kr) received the B.S. degree in computer science from KAIST, South Korea, in 2010, and the M.S. degree in software engineering from Carnegie Mellon University, United States, in 2012. He is currently working as a senior researcher with the Center for Military Analysis and Planning, KIDA, South Korea. 
His research interests lie in practical, reliable AI and its application to defense systems, as well as data-based software engineering, software architecture based testing.
}
\end{IEEEbiography}
\vspace{-1cm}
\begin{IEEEbiography}[{\includegraphics[width=1in,height=1.25in,clip,keepaspectratio]{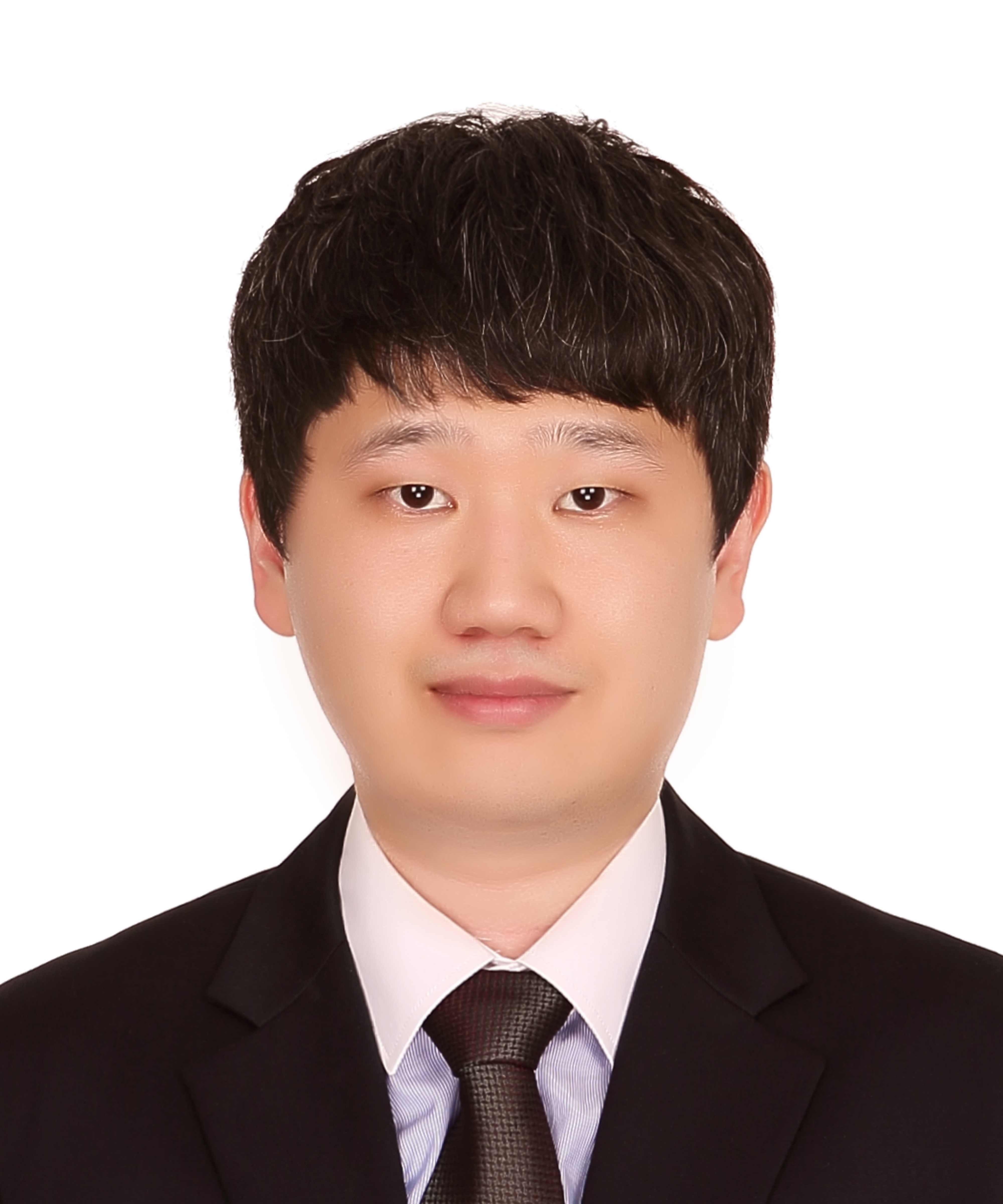}}]{Yeongjoon Kang} \textcolor{black}{(yjkang@kida.re.kr) received the B.S. and M.S. degrees in mathematical science from KAIST, South Korea, in 2016, and 2018, respectively. He is currently working as a researcher with the Center for Military Analysis and Planning, KIDA, South Korea. 
His research interests include wargame and discrete event modeling and simulation.
}
\end{IEEEbiography}
\vspace{-1cm}
\begin{IEEEbiography}[{\includegraphics[width=1in,height=1.25in,clip,keepaspectratio]{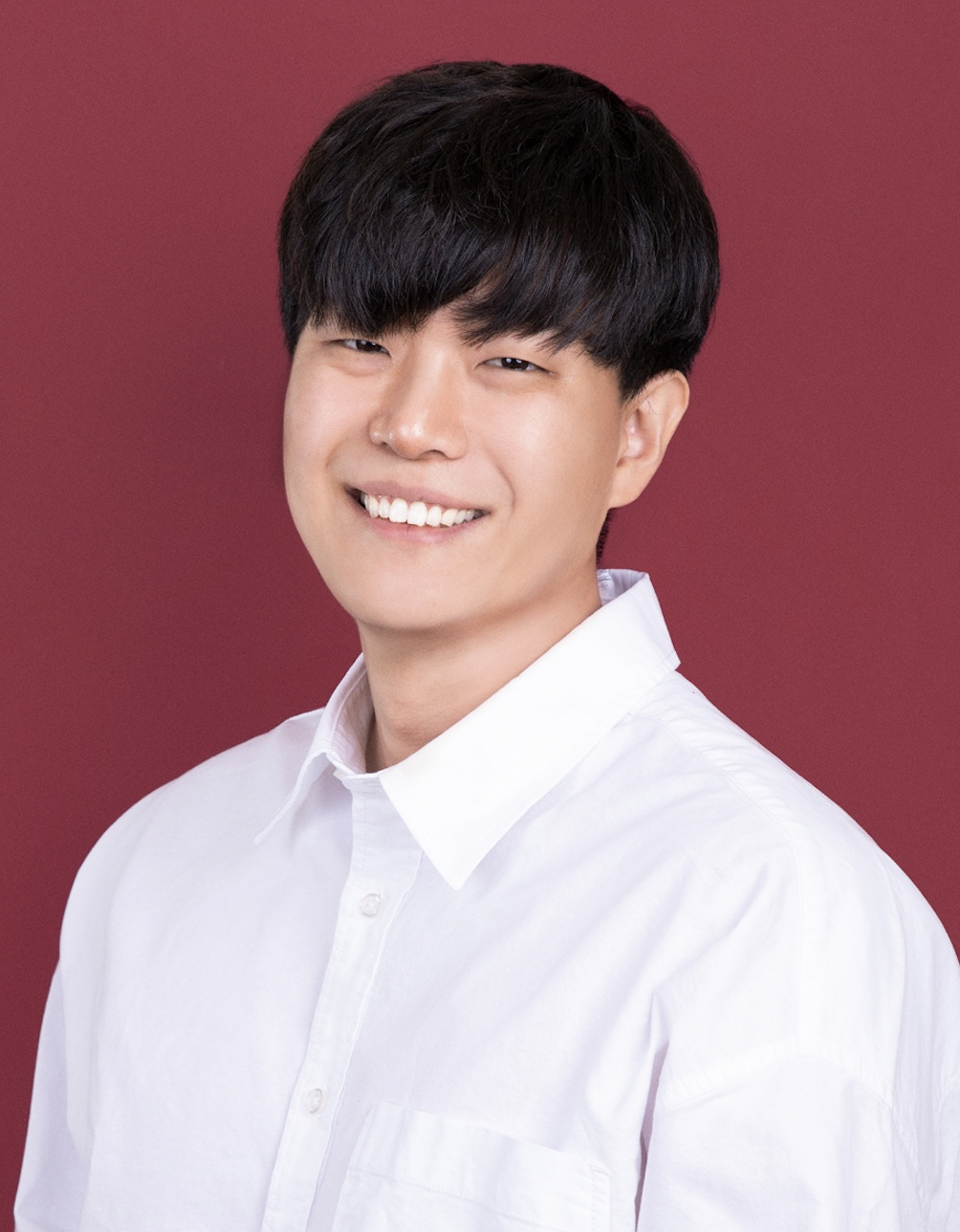}}]{Jonghoe Kim} \textcolor{black}{(kimjh@kida.re.kr) received the B.S., M.S. and Ph.D. degrees in industrial and systems engineering from KAIST, South Korea, in 2008, 2010, and 2014, respectively. He is currently working as a research fellow with the Center for Military Analysis and Planning, KIDA, South Korea. 
His research interests include vehicle routing problem, concerted system design, real-time system control, and discrete event simulation.
}
\end{IEEEbiography}
\vspace{-1cm}
\begin{IEEEbiography}[{\includegraphics[width=1in,height=1.25in,clip,keepaspectratio]{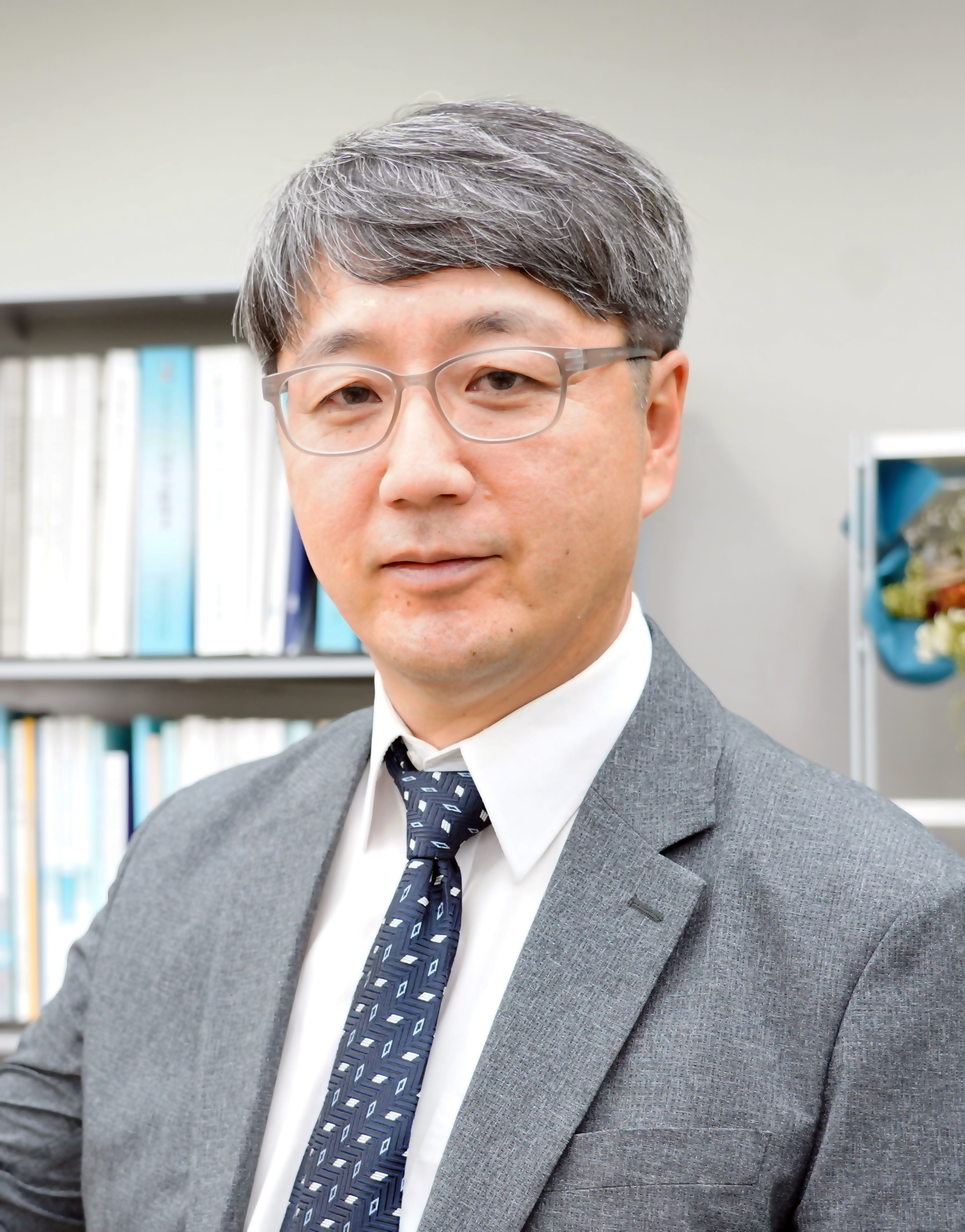}}]{Joonhyuk Kang} \textcolor{black}{(jkang@kaist.ac.kr) received the B.S.E. and M.S.E. degrees from Seoul National University, Seoul, South Korea, in 1991 and 1993, respectively. He earned his Ph.D. degree in electrical and computer engineering from the University of Texas at Austin in 2002. He is currently working as a faculty member of School of Electrical Engineering at KAIST in Daejeon, South Korea.
From 1993 to 1998, he was a research staff at SAMSUNG Electronics (Suwon, Korea), where he involved in the development of DSP-based real-time control systems. In 2000, he worked at Cwill Telecommunications (Austin, Texas, U.S.A.).
He was a visiting scholar in the School of Engineering and Applied Sciences at Harvard University (Cambridge, Massachusetts, U.S.A.) from 2008 to 2009.
His research interest includes signal processing for information transmission, security, and machine cognition.
He is a recipient of the IEEE VTS Jack Neubauer Memorial Award in 2021 for his paper titled “Mobile Edge Computing via a UAV-Mounted Cloudlet: Optimization of Bit Allocation and Path Planning”.
}
\end{IEEEbiography}

\end{document}